\documentclass[aps,prl,twocolumn,showpacs]{revtex4}

\newcommand{\Jeff}{J_{\mbox{\footnotesize eff}}}

\usepackage{graphicx}

\begin{document}

\title{Coherent control of interacting particles using dynamical
and Aharonov-Bohm phases}

\author{C.E.~Creffield}
\affiliation{Dpto de F\'isica de Materiales, Universidad
Complutense de Madrid, E-28040, Spain}

\author{G.~Platero}
\affiliation{Instituto de Ciencia de Materiales, CSIC,
Cantoblanco, Madrid, E-28049, Spain}

\date{\today}


\begin{abstract}
A powerful method of manipulating the dynamics of quantum coherent 
particles is to control the phase of their tunneling. 
We consider a system of two electrons hopping
on a quasi one-dimensional lattice in the presence of
a uniform magnetic field, and study the effect of adding a time-periodic 
driving potential. We show that the dynamical phases produced by the
driving can combine with the Aharonov-Bohm phases to give
precise control of the localization and dynamics of the
particles, even in the presence of strong particle interactions.
\end{abstract}

\maketitle

{\em Introduction -- }
Experimental advances in producing low dimensional semiconductor nanostructures
has given the opportunity of studying quantum transport in regimes
ranging from single-particle to strongly correlated.
The excellent coherence properties of these devices, together with
the degree of control over their geometry and specifications,
makes them ideal candidates for studying {\em coherent transport}, 
where quantum interference is used to regulate
the movement of particles. Such control is particularly vital
for quantum information applications, in which the coherence and entanglement
of the initial state must be preserved during the evolution of the
system, and nanostructures represent one of the most promising
avenues for realizing quantum computing elements.

\begin{center}
\begin{figure}
\includegraphics[width=0.35\textwidth,clip=true]{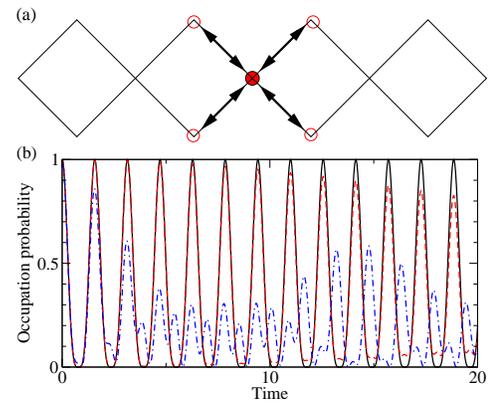}
\includegraphics[width=0.35\textwidth,clip=true]{fig1b}
\caption{(a) Schematic arrangement of the
``rhombus chain'' lattice. When each plaquette is threaded by a flux $\Phi=\pi$,
a particle initialized on the central site (filled circle) will oscillate
to its nearest neighbors (empty circles), but will not
propagate through the lattice. The circled sites form
an Aharonov-Bohm cage.
(b) Occupation probability of the central site for a two electron system.
The non-interacting case (solid black line) displays regular, non-decaying 
oscillations, indicating perfect AB-caging. For weak interaction ($U=0.1J$,
dashed red line) the oscillations decay; the caging
is no longer complete and the particle can leak away. 
For stronger interactions ($U=J$, dot-dashed blue line) 
the oscillations decay even more rapidly, and soon become 
irregular.}
\label{fig1}
\end{figure}
\end{center}

If we consider a particle hopping on a lattice, interference will occur
if the hopping acquires a phase factor. A direct way of
doing this is to apply a magnetic field, which produces the well-known
Aharonov-Bohm (AB) phase. In Ref.\onlinecite{vidal_1998} it was shown
that such phases could produce a localization effect termed 
{\em AB caging}, in which destructive interference bounds
the set of sites that can be visited by an initially localized
wavepacket. For example, in the rhombus-chain lattice shown in 
Fig.\ref{fig1}a, a particle initialized on the central site will not spread
out along the lattice if each plaquette is threaded by
a single flux quantum ($\Phi = \pi$). 
This caging effect has been observed experimentally
in superconducting wire networks \cite{expt_wires},
mesoscopic semiconductor lattices \cite{expt_meso}, and
arrays of Josephson junctions \cite{expt_jj}. The Rashba effect can 
similarly be used to generate tunneling phases, and it has
been proposed \cite{rashba} to use the spin-orbit coupling in
semiconductor nanostructures to produce an analogous caging effect.

A contrasting form of localization, also arising from quantum
interference, is termed \cite{hanggi}
``coherent destruction of tunneling'' (CDT). This arises in systems
subjected to a time-periodic driving field. Tunneling processes
acquire phase factors from the interaction of the
system with the driving, producing an effective renormalization
of the tunneling. For the most common case of sinusoidal driving,
for example, the single-particle tunneling is renormalized by 
the zeroth Bessel function ${\cal J}_0$ \cite{shirley}.

AB caging is resistant to small quantities of disorder \cite{vidal_2001},
but is swiftly destroyed by interactions (see Fig.\ref{fig1}b) due to the
formation of spatially-extended states \cite{vidal_2000}.
In this paper we consider a system of two interacting electrons,
and show that by combining a high-frequency driving potential with the 
magnetic flux, stable AB caging can be restored. This occurs when
CDT causes the (repulsive) electrons to bind
together into a composite object of charge $2e$ termed
a ``doublon'', which can then be caged by
the magnetic flux. We then go on to consider the effect of a
low frequency driving field, and show that this gives rise to an unusual
form of propagation in which the doublon moves in steps of 
two lattice sites, via the {\em virtual} occupation of the intermediate sites.

{\em Model -- } We consider a system of two electrons hopping 
on a chain of connected rhombi, as shown in Fig.\ref{fig1}a.
When propagation from site to site is coherent this system
can be modeled well \cite{cec_gp_2002} by a single-band Hubbard model
\begin{equation}
H_0 = -J \sum_{\langle j, k \rangle \sigma} \left[ e^{i \phi_{jk}}  
c^{\dagger}_{j \sigma} c^{ }_{k \sigma} + \mbox{H.c.} \right] + 
U \sum_{j} n_{j \uparrow} n_{j \downarrow} \ , 
\label{hamilton}
\end{equation}
where $J$ is the intersite tunneling between
nearest neighbors $\langle j, k \rangle$, and $U$ represents the energy
cost of doubly-occupying a lattice site. The operators
$c^{\dagger}_{j \sigma} \ / \ c^{ }_{j \sigma}$ are the usual
creation / annihilation operators for an electron
of spin $\sigma$ on site $j$, and $ n_{j \sigma}$ is the
standard number operator. Each rhombus is
threaded by a magnetic flux, $\Phi$, giving rise to the AB
phases $\exp \left[ i \phi_{jk} \right]$ on the tunneling terms.
In semiconductor quantum dots singlet-triplet mixing
terms are typically rather weak, arising from the nuclear hyperfine 
interaction \cite{hyperfine}, and accordingly we neglect
their effect here. We also confine
our attention in this work to the singlet subspace, where
one electron is spin-up and the other is spin-down,
since in the triplet subspace the Hubbard-interaction
is not operative.

{\em Results -- } In Fig.\ref{fig1}b we show the time-evolution
of the system's wavefunction by plotting its overlap 
with the initial state, which consists of both electrons occupying 
the central site. In order to produce AB caging
we set the applied flux to $\Phi = \pi$ \cite{vidal_1998}.
In the absence of interactions ($U=0$), the overlap
displays regular sinusoidal oscillations, indicating that the pair of electrons
periodically revert to their original configuration. Examining
their dynamics in detail reveals that the electrons periodically oscillate
from the initial site to its nearest neighbors, but propagate no
further down the lattice due to the caging effect.
Raising the interaction strength to $U=0.1J$ causes the AB-cage to partially
open, allowing the electrons to spread over the whole lattice, and 
so causing the oscillations in the overlap to decay.
As $U$ is increased further this leakage occurs more rapidly.

We now consider adding a periodically-oscillating potential
that rises linearly along the lattice
\begin{equation}
H(t) = H_0 + E \sin \omega t \sum_j 
x_j \left( n_{j \uparrow} + n_{j \downarrow} \right)  \ .
\end{equation}
Here $E$ and $\omega$ parameterize the amplitude and frequency of
the potential respectively, and $x_j$ is the x-component of the 
location of site $j$.
As $H(t)$ is periodic in time, $H(t) = H(t + T)$ where
$T = 2 \pi/\omega$ is the period of the driving, the Floquet
theorem allows us to write solutions of the Schr\"odinger equation
in the form
$| \Psi(t) \rangle =  e^{-i \epsilon_n t} \ | \psi_n(t) \rangle $.
Here $\epsilon_n$ are the Floquet quasienergies, and
$| \psi_n(t) \rangle$ are a set of $T$-periodic
functions termed Floquet states. The quasienergies represent
the appropriate generalization of energy eigenvalues to the
case of a periodically-driven system, and have an analogous
role in elucidating the system's behavior.
In particular, exact 
or near-degeneracy of quasienergies results
in the suppression of tunneling between the associated
Floquet states: the phenomenon known as CDT \cite{hanggi}.

\begin{center}
\begin{figure}
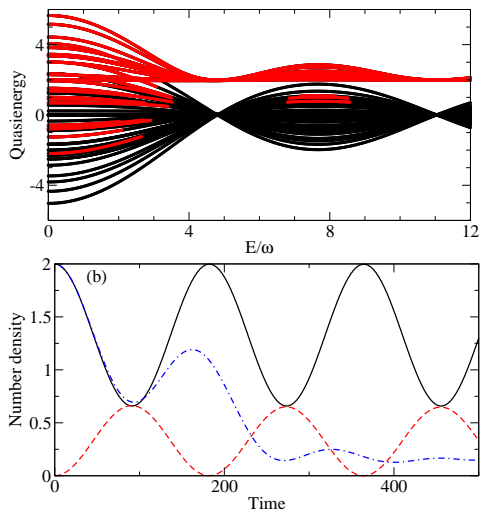

\includegraphics[width=0.35\textwidth,clip=true]{fig2a}
\includegraphics[width=0.35\textwidth,clip=true]{fig2b}
\caption{(a) Floquet quasienergies for a periodically-driven
two electron system. Parameters are $U=2 J, \ \omega=16 J$
(the high-frequency regime). The quasienergies fall into two
minibands: the lower (black) corresponds to Floquet states in which the
electrons are separated, the higher (red) to states in which both 
electrons occupy
the same site. The minibands are separated by the Hubbard gap $U$.
Both minibands collapse at $E / \omega = 4.80, \ 11.04$, corresponding to CDT.
(b) For $E/\omega = 4.80$ and $\Phi=\pi/2$ the occupation of the central
site (solid black line) oscillates as electrons periodically tunnel
to and from their nearest neighbors (see Fig.\ref{fig1}a).
For ease of comparison, the occupation of the neighboring site (dashed red line)
is multiplied by 2. When $\Phi$ is tuned away from this value AB-caging
does not occur; the dot-dashed blue line shows the occupation of the central 
site for $\Phi=0$ for which the electrons diffuse throughout the lattice.}
\label{fig2}
\end{figure}
\end{center}

{\em High frequency ($\omega > U$) -- }
In Fig.\ref{fig2}a we show the Floquet quasienergies for a system
in the high-frequency regime, where $\omega$ is the dominant energy scale.
We see that the quasienergies fall into two minibands; the higher miniband
corresponding to Floquet states in which sites are doubly occupied, and the
lower to states in which the electrons are separated in space.
Both bands ``collapse'' at specific values of $E/\omega$.
In the high-frequency regime, a perturbative calculation
of the Floquet system \cite{holthaus,cec_gp_2002} reveals that
the driving potential has the effect of renormalizing the
tunneling to an effective value
$J \rightarrow \Jeff = J {\cal J}_0(E x/\omega)$,
where $x = 1/2$ is the x-component of the spacing between neighboring
lattice sites in units of the lattice spacing.
Thus when $E / \omega \simeq 4.80, 11.04 \dots$, which 
correspond to zeros of ${\cal J}_0$,
the effective tunneling is suppressed and the miniband collapses. 

The driving field thus provides a handle to directly control 
the ratio $U / \Jeff$, which can be
enhanced by tuning the driving parameters near to a zero of
${\cal J}_0$. This has been used, for example, to induce the
Mott transition in cold atom systems \cite{pisa_mott}.
Another consequence of enhancing this ratio is the creation of 
{\em repulsively-bound} pairs \cite{cec_gp_array,zoller}, or ``doublons''.
Since binding is usually associated with attractive forces, it may 
seem counter-intuitive
for strong repulsive interactions to also produce this effect.
Qualitatively it can be understood from an energetics argument. The
energies of a one dimensional lattice form a Bloch band with a width of 
$2 J$, and thus
the maximum kinetic energy carried by two free particles
is $4 J$. If the particles are initially prepared in a state
with a potential energy much greater than $4 J$, the initial 
state then cannot decay
without the mediation of dissipative processes. In sufficiently clean systems
the only decay path \cite{decay} is via high-order scattering processes with
unbound particles that carry off the excess energy, resulting in a doublon lifetime
that depends exponentially on $U / \Jeff$.

To observe the dynamics of doublons we require $\Jeff$ to be close to
zero (to enhance the ratio $U/\Jeff$ for doublon formation to occur),
but to be sufficiently large for the doublon tunneling to be non-negligible. 
In Fig.\ref{fig2}b we show the occupation of the central site and its nearest neighbors
for driving parameters $E /\omega = 4.80$. For general values of the magnetic flux $\Phi$ 
the doublon slowly diffuses through the lattice, the two electrons remaining bound together,
but not being localized. For the case of $\Phi = \pi/2$, however,
the dynamics again shows a regular oscillatory motion due to AB-caging.
Thus even though the system is strongly interacting ($U=2 J$)
AB-caging {\em can} nonetheless be induced. Since the doublon 
has a charge of $2e$, however,
the caging now occurs when $\Phi$ is equal to {\em half} a flux quantum.

{\em Resonant frequency --}
We now consider raising the interaction strength further. When $\omega$ is 
no longer the dominant energy scale, the simple perturbation theory must be
generalized to include interactions \cite{cec_gp_2002, cec_gp_sq}.
This reveals that in this regime the effect of the driving field is particularly
strong when it is resonant with the driving frequency $U = n \omega$, where
$n$ is an integer. Away from these specific frequencies the driving 
produces little structure in the quasienergy spectrum.

\begin{center}
\begin{figure}
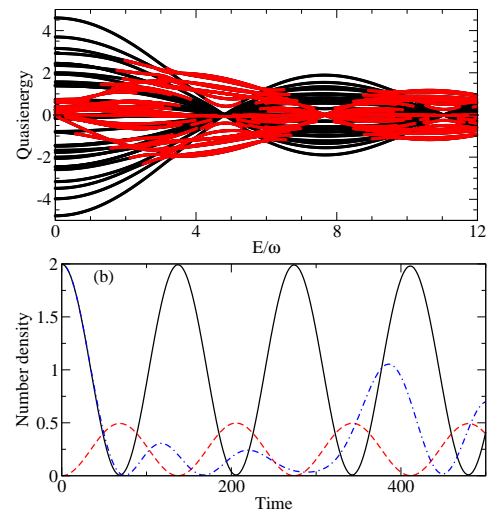

\includegraphics[width=0.35\textwidth,clip=true]{fig3a}
\includegraphics[width=0.35\textwidth,clip=true]{fig3b}
\caption{(a) Resonant driving, $U=16 J, \ \omega=16 J$. 
Quasienergies corresponding to states in which the
electrons are separated (black) are modulated by the zeroth Bessel function
${\cal J}_0(E/2 \omega)$. The quasienergies corresponding
to doubly-occupied states (red) are instead modulated by ${\cal J}_1(E / 2 \omega)$,
and show band collapses at $E x/\omega \simeq 7.66, \ 15.44$. 
(b) For $E /\omega = 7.50$ and $\Phi=\pi/2$  AB-caging occurs,
and the occupation of the central site (solid black line) 
oscillates from 2 to zero, while correspondingly the
occupation of the neighboring sites (dashed red line) oscillates 
between zero and 0.5.
Tuning $\Phi$ away from this value destroys the AB-caging; the dot-dashed 
blue line shows the occupation of the central 
site for $\Phi=0$.}
\label{fig3}
\end{figure}
\end{center}

In Fig.\ref{fig3}a we show  the quasienergies obtained for parameters
$U=16J$ and $\omega=16J$: the $n=1$ resonance. As before the Floquet
states in which the electrons are separated fall into a miniband modulated 
by ${\cal J}_0(E / 2 \omega)$. As predicted by perturbation 
theory \cite{cec_gp_array},
however, the quasienergies in which the Hubbard interaction is operative
are instead modulated by ${\cal J}_n(E/ 2 \omega)$ where $n$ is the order
of the resonance. Consequently doublon formation now occurs
when $E / 2 \omega$ is tuned near to a zero of ${\cal J}_1$.
This is shown in Fig.\ref{fig3}b; 
when $E /\omega = 7.60$, close to the first zero of ${\cal J}_1$
a stable doublon is formed, which undergoes AB-caging
when the applied  flux is set to $\Phi=\pi/2$. As before
AB-caging only occurs for this specific flux value, and for other
values of $\Phi$ the doublon simply spreads through the lattice.
It is interesting to note that unlike the high-frequency case, the
oscillation is now complete; the central site empties completely as
the doublon tunnels to the other sites in the AB-cage. This is a consequence
of the resonance condition. In an effect analogous to photon assisted tunneling,
the doublon can absorb energy from the driving field
to exactly compensate for $U$, meaning that
the system becomes effectively non-interacting. 

\begin{center}
\begin{figure}
\includegraphics[width=0.49\textwidth,clip=true]{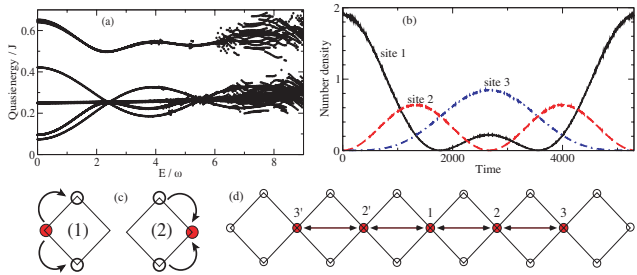}
\caption{(a) For low driving frequencies,
$U=16 J, \ \omega=2J$, the quasienergies corresponding
to doubly-occupied states form a miniband
modulated by ${\cal J}_0(E /\omega)$.
(b) Doublon dynamics for a 19-site system, 
where $E /\omega=2.30$ to suppress the direct doublon-tunneling.
The doublon is initialized on the central site 1 (see below) 
and propagates along the spine of the lattice. After reaching 
site 3 (3') it is reflected, and returns to the center. 
(c) When the doublon tunneling is suppressed, motion occurs
via second order tunneling; the doublon first unbinds (1) and
then recombines (2). The doublon thus propagates 
in steps of {\em two} lattice sites,
occupying the intermediate sites only virtually.
(d) Schematic arrangement of the 19-site lattice. Filled
circles indicate ``spinal'' sites, empty circles the ``edge'' sites.
The second order tunneling clearly does not connect spinal sites to
edge sites.}
\label{fig4}
\end{figure}
\end{center}

{\em Low frequency ($\omega < U$) -- }
Finally we consider the case when the driving frequency is much smaller
than the interaction strength. If $U/J$ is sufficiently large then
doublons will form in the static system, and 
the driving field can now be used to control their motion.
In Fig.\ref{fig4}a we show the Floquet spectrum
for states in which sites are doubly-occupied, which we can regard
as a miniband of doublon states. This miniband is clearly modulated
by ${\cal J}_0(2 E x/\omega)$, where the factor of $2$ in the argument of 
the Bessel function occurs because the doublon has charge $2e$.
The miniband structure persists until $E / \omega \geq 8$,
which marks the onset of resonant driving \cite{cec_gp_array}. 
For these parameters this will
be an $n=8$ resonance (since $U/\omega = 8$), 
but due to the strong driving potential required to
reach this regime, we do not study it further here.

Tuning $E / \omega$ close to 2.40 thus has the effect of suppressing
direct tunneling of the doublon. Motion is still possible, however,
via second order tunneling processes; the bound electrons hop to
intermediate lattice sites (incurring an energy cost of $U$), and 
then recombine after another hopping process, as illustrated schematically 
in Fig.\ref{fig4}c. 
This motion has the unusual feature that it occurs in steps of two lattice 
sites, with the intermediate sites being occupied only virtually.
If we separate the rhombus-chain lattice into ``edge'' and
''spinal'' sites, as shown in Fig.\ref{fig4}d, it can be easily seen
that this tunneling process does not connect spinal sites to
edge sites, and so produces a separation between these two sub-lattices.

In Fig.\ref{fig4}b we show the result of initializing the system
in a doublon state at the center of a 19-site lattice.
We can see that the doublon propagates smoothly along the
spine of the lattice, the occupation of edge sites
always remaining less than $0.006$.
This propagation occurs symmetrically to the left and right,
and so behaves as an electronic beam-splitter \cite{splitter}.
On reaching the final site (site 3, as shown in Fig.\ref{fig4}d)
the wavepacket is reflected and revives in the central site
with excellent fidelity. We note that the doublon does not
reach site $4$, although in principle the
second-order tunneling process should permit this.
This is essentially a finite-size effect \cite{holthaus};
the states that project onto the terminating sites ($4$ and $4'$) 
have a different symmetry to the other spinal states, and so
lie in a different miniband, thereby isolating them from the 
dynamics of the spinal states.

{\em Conclusions -- } Combining AB-phases
with a time dependent driving potential gives a extremely
rich behavior. In particular, we have shown that AB-trapping
can occur in an interacting system, by using CDT to convert
pairs of electrons into doublons.
The resonant behavior displayed by the driven system also provides
a convenient means to measure the strength of interactions
by suitably tuning the value of $\omega$.
Reducing the frequency of the driving  gives complete coherent
control over the dynamics of the doublons, allowing them to
be localized within an AB-cage, or to
propagate via an unusual second-order tunneling process.
This permits the creation and control of spatially-separated
entangled states of two electrons via the beam-splitter effect,
with many potential applications to quantum information.
The fine discrimination of this
phase-control method also allows doublons and unpaired electrons 
to be separately controlled, allowing in principle the coherent
spatial separation of these components, 
and opening up exciting possibilities for quantum transport
in interacting fermion systems.

\bigskip

This research was suported by by the Ram\'on y Cajal program (CEC), and by
Spanish MICINN through grants FIS-2007-65723 (CEC) and
MAT2008-02626 (GP).

\end{document}